\documentclass[12pt]{article}

\renewcommand{\thefootnote}{\fnsymbol{footnote}}

\usepackage{amsbsy,amssymb,latexsym,amsfonts,amsmath}
\usepackage{mathrsfs}
\usepackage{graphicx}
\usepackage{bm}
\usepackage{here}
\usepackage{comment}
\usepackage{amsmath}
\usepackage{cases}
\usepackage {empheq}
\usepackage{color}
\usepackage{multirow}
\usepackage{comment}
\makeatletter
\def\EqNumText{\refstepcounter{equation}\cdots\tagform@\theequation}%
\makeatother

\newcommand{\bel}[1]{\begin{equation}\label{#1}}
\newcommand{\bal}[1]{\begin{eqnarray}\label{#1}}
\newcommand{\be}{\begin{equation}}
\newcommand{\ee}{\end{equation}}

\begin{document}
%
%
\begin{titlepage}
\begin{flushright}
\normalsize
~~~~
NITEP 100\\
OCU-PHYS 538\\
June 20, 2021 \\
\end{flushright}

\vspace{15pt}

\begin{center}
{\LARGE  Gauge Symmetry Enhancement by Wilson Lines   \\
in Twisted Compactification} \\
\end{center}

\vspace{23pt}

\begin{center}
{ H. Itoyama$^{a, b,c}$\footnote{e-mail: itoyama@osaka-cu.ac.jp}, Yuichi Koga$^b$\footnote{e-mail: k-yuichi@yj.osaka-cu.ac.jp}
 and   Sota Nakajima$^b$\footnote{e-mail: nakajima@rx.osaka-cu.ac.jp}   }\\

%
\vspace{10pt}
%

$^a$\it Nambu Yoichiro Institute of Theoretical and Experimental Physics (NITEP),\\
Osaka City University\\
\vspace{5pt}

$^b$\it Department of Mathematics and Physics, Graduate School of Science,\\
Osaka City University\\
\vspace{5pt}

$^c$\it Osaka City University Advanced Mathematical Institute (OCAMI)

\vspace{5pt}

3-3-138, Sugimoto, Sumiyoshi-ku, Osaka, 558-8585, Japan \\

\end{center}
%
\vspace{15pt}
\begin{center}
Abstract\\
\end{center}
   We point out a simple realization of gauge symmetry enhancement by Wilson lines in QFT with
   twisted compactification in extra dimensions. We illustrate this in the field contents taken
   from heterotic supergravity in arbitrary dimensions toroidally compactified.

\vfill

\end{titlepage}

\renewcommand{\thefootnote}{\arabic{footnote}}
\setcounter{footnote}{0}


In search of proper vacua that serve for the unification of forces, we often deal with a category of quantum field theories deformed both by boundary conditions and by local as well as non-local operators.
In this letter, we point out that simultaneous consideration of the former by a $Z_2$ twist taken from the representation of the gauge group and
the latter  by a set of Wilson lines introduced in extra dimensions leads us to a perturbative realization of
landscape of extrema where gauge symmetry enhancement takes place.

Twisting operations are often adopted in order to break supersymmetry\cite{Spontaneous Breaking of Supersymmetry Through Dimensional Reduction}. In the context of string theory, various models without spacetime supersymmetry which are constructed by compactifications with twists have been found and investigated (e.g. \cite{Dixon:1986iz, Itoyama:1986ei,Nair:1986zn} for heterotic strings, \cite{Blum:1997gw} for open strings ). In particular, it is known that compactifying supersymmetric string models with certain twists can realize exponentially suppressed cosmological constants\cite{Itoyama:1986ei}. Such string models have been focused on in order to look into top-down scenarios from non-supersymmetric string theory\cite{Itoyama:2019yst,Itoyama:2020ifw,Itoyama:2021,Abel:2015oxa,Faraggi:2009xy,Kounnas:2015yrc}.

Gauge symmetry enhancement is a ubiquitous phenomenon in string theory realized at the Planck scale.
While our current work is motivated in part by the recent progress\cite{Itoyama:2019yst,Itoyama:2020ifw,Itoyama:2021} in the heterotic interpolating model\cite{Itoyama:1986ei},
we will demonstrate directly in QFT with twisted compactification without referring to string theory,
taking anomaly free field contents from ten-dimensional heterotic supergravity theory\cite{Chamseddine:1980cp,Chapline:1982ww,Green:1984sg,Candelas:1985en}.
 In fact, the one-loop vacuum energy is dominated by massless spectra with Kaluza-Klein excitations summed over, and the part with the Planck scale UV cutoff is exponentially suppressed as radius parameters get large.
\\

Let us begin with a general discussion. The field contents we pay attention to are the adjoint vectors and the adjoint spinors in ten spacetime dimensions while more general setup is certainly possible. We consider a general $D$-dimensional toroidal compactification. Without losing generality, we introduce $\boldsymbol{Z}_{2}$ twist only in the $x^{9}$-direction. Let us denote the 10-dimensional coordinates by $x^{M} = (x^{\mu}; x^{a},x^{9}), ~~(\mu = 0,\ldots,9-D;~a = 10-D,\ldots,8)$. In addition, we will often assign suffix $j$ for $j=10-D,\cdots,9$. The boundary condition for a field $\phi(x^{\mu};x^{a},x^9)$ is
 \begin{align}\label{periodicity}
 &\phi(x^{\mu};x^{a}+2\pi R^{a},x^{9})= \phi(x^{\mu}; x^{a},x^{9}),\\
 \label{twisted}
 &\phi(x^{\mu};x^{a},x^{9}+2\pi R^{9})= (-)^{F}T_{g}\phi(x^{\mu}; x^{a},x^{9}),
 \end{align}
 where $F$ is the spacetime fermion number,
 and an operator $T_{g}$ is a representation matrix of the gauge group $g$ such that $T_{g}^{2}=1$. We will obtain below the mass formulae for the adjoint vector and those for the adjoint spinor under the conditions \eqref{periodicity}, \eqref{twisted} with Wilson lines turned on.
 Let $A^{M}$ be a vector field lying in the adjoint representation. Taking the Lorentz gauge $\partial^{M}A_{M}=0$, the equation of motion reads
 \begin{align}\label{eom for vector}
   0=[D^{M},F_{MN}]=\partial^{M}\partial_{M}A_{N}+2i[A_{M},\partial^{M}A_{N}]-i[A^{M},\partial_{N}A_{M}]-[A^{M},[A_{M},A_{N}]].
 \end{align}
 Here, $F^{MN}=\frac{1}{i}[D^{M},D^{N}],~D^{M}=\partial^{M}+iA^{M}$. The compactified components $A^{j}$ receive VEV's through the Wilson line operators and we write as
 \begin{align}\label{WL}
  A^{j}=\mathcal{A}^{j}+\tilde{A}^{j} ,~~~\mathcal{A}^{j} = \sum_{I=1}^{r} \mathcal{A}^{j,I} H^{I},
 \end{align}
 where $r=rank(g)$, the constants $\mathcal{A}^{j,I}$ are from the Wilson lines and $H^{I}$ are the Cartan generators.
From the variation of the action by $\tilde{A}^{M}=\left( A^{\mu},\tilde{A}^{j}\right) $, we obtain
 \begin{align}\label{eom2 for vector}
   0&=\partial^{\nu}\partial_{\nu}\tilde{A}_{M}+2i[A_{\nu},\partial^{\nu}\tilde{A}_{M}]-i[A^{\nu},\partial_{M}A_{\nu}]-[A^{\nu},[A_{\nu},\tilde{A}_{M}]]\nonumber\\
   &~~+\partial^{i}\partial_{i}\tilde{A}_{M}+2i[A_{i},\partial^{i}\tilde{A}_{M}]-[A^{i},[A_{i},\tilde{A}_{M}]].
 \end{align}
 The second line with $A^{i}$ replaced by $\mathcal{A}^{i}$ contributes to the mass term which we evaluate below. In the Cartan-Weyl basis, the twisted boundary condition \eqref{twisted} is
 \begin{align}\nonumber
 A^{M}(x^{\mu};x^{a},x^{9}+2\pi R^{9})&= UA^{M}(x^{\mu}; x^{a},x^{9})U^{\dagger}\\
 =& \sum_{I=1}^{r}A^{M,I}(x^{\mu}; x^{a},x^{9})H^{I}+
 \sum_{\alpha\in \Delta'}e^{2\pi ic\cdot\alpha} A^{M,\alpha}(x^{\mu}; x^{a},x^{9})E^{\alpha},
 \end{align}
 where $U=e^{2\pi i c\cdot H}$ and $c_{I}$ are the components of a dual vector such that $2c_{I}$ are those lying in the dual of the root lattice, and $\Delta'$ denotes the set of nonzero roots of the gauge group. The dot $\cdot$ denotes the pairing. The Kaluza-Klein expansion of the adjoint vectors is
 \begin{equation} \label{KK for vector}
 \begin{split}
 A^{M,I}(x^{\mu};x^{a},x^{9})
 =&\sum_{\boldsymbol{m}\in \boldsymbol{Z}^D} A_{\boldsymbol{m}}^{M,I}(x^{\mu})e^{i\sum_{j} \left( \frac{m_{j}x^{j}}{R^{j}}\right) },\\
 A^{M,\alpha}(x^{\mu};x^{a},x^{9})=&
 \sum_{\boldsymbol{m}\in \boldsymbol{Z}^D} A_{\boldsymbol{m}}^{M,\alpha}(x^{\mu})
 e^{i\sum_{a} \left( \frac{m_{a}x^{a}}{R^{a}}\right) +i\frac{(m_{9}+c\cdot \alpha)x^{9}}{R^{9}}} ,
 \end{split}
 \end{equation}
 where $\boldsymbol{m}$ denotes $(m_{a},m_9)$.
 We read off the mass formulae for $A_{\boldsymbol{m}}^{M,I}(x^{\mu})$ and for $A_{\boldsymbol{m}}^{M,\alpha}(x^{\mu})$ with the Wilson lines $t^{j,I} \equiv R^{j}\mathcal{A}^{j,I}$ ($j$ not summed) from \eqref{eom2 for vector} and \eqref{KK for vector}:
 \begin{align}\label{mass for vector}
   M_{B,I,\boldsymbol{m}}^{2}=\sum_{j=10-D}^{9}\left( \frac{m_{j}}{R^{j}}\right)^2 ,~~~
   M_{B,\alpha,\boldsymbol{m}}^{2}=\sum_{a=10-D}^{8}\left( \frac{m_{a}+t^{a}\cdot \alpha}{R^{a}}\right)^2 + \left( \frac{m_{9}+c \cdot \alpha + t^{9}\cdot \alpha}{R^{9}}\right)^2.
 \end{align}
 Note that the states with all $m_{j}=0$ in the direction of the Cartan subalgebra are massless, and the rank of the gauge group stays the same upon compactification.\\

 As for the adjoint spinor $\psi$, the equation of motion is
 \begin{align}\label{eom for spinor}
   0=\Gamma^{M}D_{M}\psi=\Gamma^{\mu}D_{\mu}\psi + \Gamma^{j}D_{j}\psi,
 \end{align}
 where $\Gamma^{M}$ satisfy $\{ \Gamma^{M}, \Gamma^{N} \}=-2\eta^{MN}$, $\eta^{MN} = diag(-1,+1,\ldots +1)$ and $D_{M}\psi=\partial_{M}\psi+i[A_{M},\psi]$.
 The second term of the right-hand side in \eqref{eom for spinor} is a mass term. The twisted boundary condition \eqref{twisted} reads
 \begin{align}\nonumber
 \psi(x^{\mu};x^{a},x^{9}+2\pi R^{9})&= -U\psi(x^{\mu};x^{a},x^{9}) U^{\dagger}\\
 &=-\left( \sum_{I=1}^{r}\psi^{I}(x^{\mu};x^{a},x^{9})H^{I}+\sum_{\alpha\in \Delta'}e^{2\pi ic\cdot\alpha} \psi^{\alpha}(x^{\mu};x^{a},x^{9})E^{\alpha}\right).
 \end{align}
 The Kaluza-Klein expansion is
 \begin{equation} \label{KK for spinor}
 \begin{split}
 \psi^{M,I}(x^{\mu};x^{a},x^{9})
 =&\sum_{\boldsymbol{m}\in \boldsymbol{Z}^D} \psi_{\boldsymbol{m}}^{M,I}(x^{\mu})e^{i\sum_{a}\left( \frac{m_{a}x^{a}}{R^{a}}\right) +i\frac{\left( m_{9}+1/2\right) x^{9}}{R^{9}}},\\
 \psi^{M,\alpha}(x^{\mu};x^{a},x^{9})=&
 \sum_{\boldsymbol{m}\in \boldsymbol{Z}^D} \psi_{\boldsymbol{m}}^{M,\alpha}(x^{\mu})
 e^{i\sum_{a}\left(  \frac{m_{a}x^{a}}{R^{a}}\right) +i\frac{(m_{9}+1/2+c\cdot \alpha)x^{9}}{R^{9}}}.
 \end{split}
 \end{equation}
 Using \eqref{eom2 for vector} and \eqref{KK for spinor}, we read off the mass formulae for $\psi_{\boldsymbol{m}}^{M,I}(x^{\mu})$ and $\psi_{\boldsymbol{m}}^{M,\alpha}(x^{\mu})$:
 \begin{equation}\label{mass for spinor}
 \begin{split}
   M_{F,I,\boldsymbol{m}}^{2}&=\sum_{a=10-D}^{8}\left( \frac{m_{a}}{R^{a}}\right)^2 + \left( \frac{m_{9}+1/2}{R^{9}}\right)^2,~\\
   M_{F,\alpha,\boldsymbol{m}}^{2}&=\sum_{a=10-D}^{8}\left( \frac{m_{a}+t^{a}\cdot \alpha}{R^{a}}\right)^2 + \left( \frac{m_{9}+1/2+c \cdot \alpha + t^{9}\cdot \alpha}{R^{9}}\right)^2.
 \end{split}
 \end{equation}
 Note that $M_{F,I,\boldsymbol{m}}$ cannot be zero whatever value $\boldsymbol{m}$ takes, and the gauginos lying in the Cartan subalgebra become massive.
 On the other hand, there are some conditions on $c \cdot \alpha$ as well as on $t^{j}\cdot \alpha$ under which $M_{F,\alpha,\boldsymbol{m}}=0$. It is convenient to define $\Delta'_{+}$ and  $\Delta'_{-}$ as
 \begin{align}
 \Delta'_{+}=\left\lbrace \left.  \alpha\in \Delta' \right|~ c\cdot\alpha\in\boldsymbol{Z} \right\rbrace,~~~~\Delta'_{-}=\left\lbrace  \alpha\in \Delta' \left|~  c\cdot\alpha\in\boldsymbol{Z}+\frac{1}{2} \right. \right\rbrace.
 \end{align}
 Then, in the absence of the Wilson lines, the mass formula \eqref{mass for vector} tells us the presence of an additional massless vector associated with $\alpha\in \Delta'_{+}$ while the mass formula \eqref{mass for spinor} tells us that of an additional massless spinor associated with $\alpha\in \Delta'_{-}$.
  The particular case $t^{a}\cdot \alpha\notin \boldsymbol{Z}$, $c\cdot\alpha=t^9\cdot \alpha=0$ has been well exploited in unified field theory model building with extra dimensions to introduce the gauge symmetry breaking\cite{Manton:1979kb,Fairlie:1979at,Hosotani:1983xw,Hosotani:1988bm,Hatanaka:1998yp,Quiros:2003}. Our general mass formulae \eqref{mass for vector}, \eqref{mass for spinor} open up a possibility to probe the landscape of perturbative vacua by non-vanishing $c\cdot\alpha$ and $t^{9}\cdot \alpha$.
   \\
   
   Let us now take a closer look at the mass formulae \eqref{mass for vector}, \eqref{mass for spinor} by a concrete case and illustrate the gauge enhancement mechanism. We take anomaly free field contents from the $E_8 \times E_8$ heterotic supergravity in ten dimensions coupled to the adjoint matter.
   Let us adopt the following twist that breaks the gauge group $E_8 \times E_8$ to $SO(16) \times SO(16)$:
   \begin{align}\label{twist}
   	c_{I}=\left( 1,(0)^7;1,(0)^7\right).
   \end{align}
   In this choice of $c_{I}$, $\Delta'_{+}$ forms the nonzero roots of $SO(16) \times SO(16)$ while $\Delta'_{-}$ gives the $\left( \boldsymbol{128},\boldsymbol{1}\right) \oplus \left( \boldsymbol{1},\boldsymbol{128}\right) $ of $SO(16) \times SO(16)$:
   \begin{align}
    \label{roots of SO(16)}
    \alpha^{I}&=\left( \underline{\pm 1, \pm 1, (0)^{6}}; (0)^{8} \right), \left( (0)^{8}; \underline{\pm 1, \pm 1, (0)^{6}} \right) \in \Delta'_{+},\\
    \alpha^{I}&=\begin{cases}
    \frac{1}{2}\left( \underline{\pm 1, \pm 1, \pm 1, \pm 1, \pm 1, \pm 1, \pm 1, \pm 1}_{+}; (0)^{8} \right)\\
    \frac{1}{2}\left( (0)^{8}; \underline{\pm 1, \pm 1, \pm 1, \pm 1, \pm 1, \pm 1, \pm 1, \pm 1}_{+} \right)
    \end{cases}\in \Delta'_{-},
   \end{align}
   where the underline represents the permutations of the components and the index $+$ ($-$) attached to the underline indicates that the number of plus signs is even (odd). We will sometimes adopt the notation to write the index $I$ dividing in half, such as $\alpha^{I}=\left( \alpha^{A},\alpha^{A'} \right)$ with $A=1, \cdots ,8$, $A'=9, \cdots ,16$.
   From the mass formulae (\ref{mass for vector}) and (\ref{mass for spinor}), we obtain the gauge bosons of $SO(16)\times SO(16)$ and the massless spinors transforming in the $\left( \boldsymbol{128},\boldsymbol{1}\right) \oplus \left( \boldsymbol{1},\boldsymbol{128}\right) $ of $SO(16)\times SO(16)$ under $t^{j,I}=0$.
 
	At generic points in the moduli space, the massless spectrum just consists of $U(1)^{16}$ gauge bosons since neither $M_{B,\alpha,\boldsymbol{m}}$ nor $M_{F,\alpha,\boldsymbol{m}}$ can take zero for generic values of $t^{j,I}$. We can however get additional massless states at special points in the moduli space that satisfy $t^{j}\cdot \alpha \in \boldsymbol{Z}$ or $t^{j}\cdot \alpha \in \boldsymbol{Z}+1/2$. We are now interested in the special directions $t^{9,I}$ in the moduli space which reflect the effect of the non-trivial boundary condition \eqref{twisted}. So, we focus on a subspace of the moduli space which satisfies $t^{a,I}=0$ so that the gauge group is $SO(16)\times SO(16)$ when $t^{9,I}=0$.
	Let us consider the following configurations of the Wilson lines:
	\begin{align}\label{condition1}
		t^{9,I}
    =\left( \left( 0\right) ^{p_{1}}, \left( 1\right) ^{p_{2}}, \left(\frac{1}{2} \right)^{q_{1}}, \left( -\frac{1}{2} \right)^{q_{2}} ; \left( 0\right) ^{p'_{1}}, \left( 1\right) ^{p'_{2}}, \left(\frac{1}{2} \right)^{q'_{1}}, \left( -\frac{1}{2} \right)^{q'_{2}}  \right),
	\end{align}
	for $p_{1} + p_{2} + q_{1} + q_{2} = p'_{1} + p'_{2} + q'_{1} + q'_{2} = 8$. First, let us focus on the nonzero roots in $\Delta'_{+}$.
	With the configuration \eqref{condition1}, a set of $\alpha\in\Delta'_{+}$ satisfying $t^{9}\cdot \alpha\in\boldsymbol{Z}$ is
	\begin{align}\label{adjoint}
		\alpha^{I}=\begin{cases}
		\left( \underline{\pm 1, \pm 1, (0)^{p-2}},(0)^{q};(0)^{8} \right),~~\left( (0)^{p}, \underline{\pm 1, \pm 1, (0)^{q-2}}; (0)^{8}\right) ,\\
		\left( (0)^{8};  \underline{\pm 1, \pm 1, (0)^{p'-2}},(0)^{q'}\right),~~\left( (0)^{8};  (0)^{p'},\underline{\pm 1, \pm 1, (0)^{q'-2}}\right),
		\end{cases}
	\end{align}
	where $p= p_1+p_2$, $q= q_1+q_2$, $p'= p'_{1}+p'_{2}$, $q'= q'_{1}+q'_{2}$. 
	This set of $\alpha$ corresponds to the nonzero roots of $SO(2p)\times SO(2q)\times SO(2p')\times SO(2q')$. Namely, turning on the Wilson lines that take the configuration \eqref{condition1}, the gauge symmetry is broken or enhanced to $SO(2p)\times SO(2q)\times SO(2p')\times SO(2q')$ from $SO(16)\times SO(16)$.
	On the other hand, the spinors associated with $\alpha\in\Delta'_{+}$ are massless if $t^{9}\cdot \alpha \in \boldsymbol{Z}+1/2$. 
	 A set of such $\alpha$ forms the weights $\left( \boldsymbol{2p}, \boldsymbol{2q}\right) $ of $SO(2p)\times SO(2q)$ and the $\left( \boldsymbol{2p'}, \boldsymbol{2q'}\right) $ of $SO(2p')\times SO(2q')$:
	 \begin{align}\label{bi-fund}
	 	\alpha^{I}=\begin{cases}
	 	\left( \underline{\pm 1,  (0)^{p-1}},\underline{\pm 1,  (0)^{q-1}};(0)^{8} \right)  , \\
	 	\left((0)^{8}; \underline{\pm 1,  (0)^{p'-1}},\underline{\pm 1,  (0)^{q'-1}} \right).
	 	\end{cases}
	 \end{align}
	If $q$ is odd, the states with $\alpha\in\Delta'_{-}$ cannot be massless, i.e. there is no $\alpha\in\Delta'_{-}$ such that $t^{9}\cdot \alpha=0$ mod 1/2. If $q$ is even, we obtain massless states with $\alpha\in\Delta'_{-}$ in addition to those with $\alpha\in\Delta'_{+}$. We shall henceforth focus on the first half components $\alpha^{A}$ of the nonzero roots\footnote{This reasoning applies the second half components $\alpha^{A'}$ as well.}. We should note that the conditions for vectors and spinors with $\alpha\in\Delta'_{-}$ to be massless are interchanged compared to that of the case with  $\alpha\in\Delta'_{+}$: $t^{9}\cdot\alpha\in\boldsymbol{Z}+1/2$ for vectors and $t^{9}\cdot\alpha\in\boldsymbol{Z}$ for spinors. Below we analyze the three cases, $q=0,2,4$,\footnote{The cases $q=6$ and the case $q=8$ are equivalent to $q=2$ and $q=0$ respectively.} and summarize the results in Table 1.	
	\begin{itemize}
    \setlength{\leftskip}{-0.5cm}
		\item $q=0$: every $\alpha\in\Delta'_{-}$ satisfies $t^{9}\cdot\alpha\in\boldsymbol{Z}$ if $p_{2} + q_{2}$ (which is just $p_{2}$ in this case) is even. Namely, the massless states are the spinors transforming in the $\boldsymbol{128}$ of $SO(16)$. 
		If $p_{2} + q_{2}$ is odd, which means that every $\alpha\in\Delta'_{-}$ satisfies $t^{9}\cdot\alpha\in\boldsymbol{Z}+1/2$, we get the massless vectors transforming in the $\boldsymbol{128}$ of $SO(16)$ in addition to those in the $\boldsymbol{120}$ of $SO(16)$. As a result, the gauge symmetry is enhanced to $E_8$ and there is no massless fermion. 

		\item $q=2$: in the case with even $p_{2} + q_{2}$, the nonzero roots in $\Delta'_{-}$ which satisfy $t^{9}\cdot\alpha\in\boldsymbol{Z}$ are
		\begin{align}\label{alpha+6,2}
			\alpha^{A}=\frac{1}{2}\left(\underline{\pm 1,\pm 1,\pm 1,\pm 1,\pm 1,\pm 1}_{+},\underline{\pm 1, \pm 1}_{+}\right),
		\end{align}
		and those which satisfy $t^{9}\cdot\alpha\in\boldsymbol{Z}+1/2$ are
		\begin{align}\label{alpha-6,2}
		\alpha^{A}=\frac{1}{2}\left(\underline{\pm 1,\pm 1,\pm 1,\pm 1,\pm 1,\pm 1}_{-},\underline{\pm 1, \pm 1}_{-}\right).
		\end{align}
		If $p_{2} + q_{2}$ is odd, $t^{9}\cdot\alpha\in\boldsymbol{Z}$ holds for \eqref{alpha-6,2} and $t^{9}\cdot\alpha\in\boldsymbol{Z}+1/2$ holds for \eqref{alpha+6,2}. The difference is only the chirality of the spinor representations  of $SO(12)\times SO(4)$, so the same massless spectra are obtained whether $p_{2} + q_{2}$ is even or odd. Noting that $\alpha^{A}$ in $\Delta'_{+}$ forms the adjoint representation of $SO(12)\times SO(4)$ for the massless vectors, the gauge symmetry is enhanced to $E_{7}\times SU(2)$. For the massless spinors, $\alpha^{A}$ in $\Delta'_{+}$ gives the bi-fundamental weight of $SO(12)\times SO(4)$. As a result, we obtain the massless spinors transforming in the $\left( \boldsymbol{56},\boldsymbol{2} \right) $ of $E_7 \times SU(2)$.

		\item $q=4$: for even $p_{2} + q_{2}$, the nonzero roots in $\Delta'_{-}$ with satisfying $t^{9}\cdot\alpha\in\boldsymbol{Z}$ are
		\begin{align}\label{alpha+4,4}
		\alpha^{A}=\frac{1}{2}\left(\underline{\pm 1,\pm 1,\pm 1,\pm 1}_{+},\underline{\pm 1, \pm 1,\pm 1,\pm 1}_{+}\right),
		\end{align}
		and with satisfying $t^{9}\cdot\alpha\in\boldsymbol{Z}+1/2$ are
		\begin{align}\label{alpha-4,4}
		\alpha^{A}=\frac{1}{2}\left(\underline{\pm 1,\pm 1,\pm 1,\pm 1}_{-},\underline{\pm 1, \pm 1,\pm 1,\pm 1}_{-}\right).
		\end{align}
		As in the $q=2$ case, $t^{9}\cdot\alpha\in\boldsymbol{Z}$ and $t^{9}\cdot\alpha\in\boldsymbol{Z}+1/2$ hold for \eqref{alpha-4,4} and \eqref{alpha+4,4} respectively for odd $p_{2} + q_{2}$ and we obtain the same spectrum in both cases.
		 Considering \eqref{adjoint} and \eqref{bi-fund} with $q=4$, the massless spectrum consists of the gauge bosons of $SO(16)$ and the massless spinors transforming in the $\boldsymbol{128}$ of $SO(16)$.

	\end{itemize}

\begin{table}[t]
	\centering
	\begin{tabular}{|c||c|c|c|} \hline
		Special points & $p_2+q_2$ & Gauge group & Massless spinors \\ \hline
		\multirow{2}{*}{$q=0,8$} & even & $SO(16)$ & $\boldsymbol{128}$ \\
		& odd & $E_8$ & $nothing$ \\ \hline
		$q=2,6$ & even/odd & $E_7 \times SU(2)$ & $(\boldsymbol{56}, \boldsymbol{2})$\\ \hline
		$q=4$ & even/odd & $SO(16)$ & $\boldsymbol{128}$ \\ \hline
	\end{tabular}
	\caption{Special points in half of the moduli space and the massless spectra.}
\end{table}
We should note that Table 1 shows only the pattern of gauge symmetry which is realized by the first half components of the Wilson line. The massless spectrum which is realized by the full set of Wilson lines is presented by the double of that shown in Table 1.
\\

 In the remainder of this letter, we discuss a relationship between the gauge symmetry enhancement and the one-loop cosmological constant.
 The $(10-D)$-dimensional cosmological constant is
 \begin{align}\label{vacuum energy}
   \Lambda^{(10-D)} &=\frac{1}{2} \sum_{\mathcal{I}}(-1)^{F} \int \frac{d^{10-D} p}{(2 \pi)^{10-D}} \log \left(p^{2}+M_{\mathcal{I}}^{2}\right) \nonumber\\
   &=-\frac{1}{2} \frac{1}{(2 \pi)^{10-D}} \sum_{\mathcal{I}}(-1)^{F} \int_{\epsilon^2}^{\infty} \frac{d \tau}{\tau^{1+(10-D)/ 2}} e^{-\pi \tau M_{\mathcal{I}}^{2} }\nonumber\\
   &=-\frac{1}{2} \frac{1}{(2 \pi)^{10-D}}\int_{\epsilon^2}^{\infty} \frac{d \tau}{\tau^{1+(10-D)/ 2}} \nonumber\\
   &~~~\times 8\sum_{\boldsymbol{m}\in \boldsymbol{Z}^{D}}\left(  \sum_{I} e^{-\pi \tau M_{B,I,\boldsymbol{m}}^{2} }
   +\sum_{\alpha\in \Delta'} e^{-\pi \tau M_{B,\alpha,\boldsymbol{m}}^{2} }
   -\sum_{I} e^{-\pi \tau M_{F,I,\boldsymbol{m}}^{2} }
   -\sum_{\alpha\in \Delta'} e^{-\pi \tau M_{F,\alpha,\boldsymbol{m}}^{2} }\right)  .
 \end{align}
where the summation of $\mathcal{I}$ is over all species of physical particles that are labelled by the internal degrees of freedom, and $\epsilon$ is a UV cutoff. The prefactor 8 implies the number of polarizations for a vector (or a spinor) in ten dimensions and the contribution from a ten-dimensional gravity multiplet is omitted in \eqref{vacuum energy}.
In our setup with mass formulae given by \eqref{mass for vector} and \eqref{mass for spinor}, the integrand of \eqref{vacuum energy} is
\begin{align}\label{integrand1}
	&\sum_{\boldsymbol{m}\in\boldsymbol{Z}^D}\Biggl\lbrace r e^{-\pi\tau \sum_{a}\left( \frac{m_{a}}{R^{a}}\right)^2 } \left( e^{-\pi\tau \left( \frac{m_{9}}{R^{9}}\right)^2}- e^{-\pi\tau \left( \frac{m_{9}+1/2}{R^{9}}\right)^2}\right)  \nonumber\\
	&~~~~~~
	+\sum_{\alpha\in \Delta'}e^{ -\pi\tau \sum_{a}\left( \frac{m_{a}+t^{a}\cdot\alpha}{R^{a}}\right)^2 }
	\left( e^{-\pi\tau \left( \frac{m_{9}+\left(c+t^{9} \right)\cdot\alpha }{R^{9}}\right)^2 }
	- e^{-\pi\tau \left( \frac{m_{9}+1/2+ \left(c+t^{9} \right)\cdot\alpha}{R^{9}}\right)^2}\right) 
	 \Biggr\rbrace.
\end{align}
By using the Poisson resummation formula
\begin{align}
	\sum_{n=-\infty}^{\infty} e^{-a (n+x)^{2}}=\sqrt{\frac{\pi}{a}}\sum_{n=-\infty}^{\infty} e^{-\pi^2 n^{2}/a + 2\pi i nx},
\end{align}
 the integrand \eqref{integrand1} gets further converted into
\begin{align}\label{integrand2}
&\tau^{-\frac{D}{2}}\left( \prod_{j}R^{j}\right) \sum_{\boldsymbol{m}\in\boldsymbol{Z}^D}
 e^{-\frac{\pi}{\tau}\sum_{j}\left( m_{j}R^{j}\right)^2} \left( 1-e^{-\pi i m_{9} }\right)  
 \left(r+ \sum_{\alpha\in\Delta'}e^{2\pi i\sum_{a}\left( m_{a}t^{a}\cdot\alpha\right)}e^{2\pi i m_{9}\left( c+t^{9}\right) \cdot\alpha}\right).
\end{align}
The factor $\left( 1-e^{-\pi i m_{9} }\right) $ implies that the contributions from even $m_{9}$ are canceled.
As a result, we obtain the following form of the cosmological constant as a function of the moduli:
\begin{align}\label{potential result}
	\Lambda^{(10-D)}&=-\frac{1}{2} \frac{1}{(2 \pi)^{10}} \left(  \prod_{j}2\pi R^{j}\right) 
	\int_{\epsilon^2}^{\infty} \frac{d \tau}{\tau^{6}}\sum_{\boldsymbol{m}\in\boldsymbol{Z}^D}
	e^{-\frac{\pi}{\tau}\sum_{a}\left( m_{a}R^{a}\right)^2} e^{-\frac{\pi}{\tau}\left( \left( 2m_{9}-1\right) R^{9}\right)^2} \nonumber\\
	&~~~~~~~~~~~~~~~~~~~~~~~~~~~
	\times 8\left(8+r+ \sum_{\alpha\in\Delta'}e^{2\pi i\sum_{a}\left( m_{a}t^{a}\cdot\alpha\right)}e^{2\pi i \left( 2m_{9}-1\right)\left( c+t^{9}\right) \cdot\alpha}\right)\nonumber\\
	& =-\frac{1}{2} \frac{1}{(2 \pi)^{10}} \left(  \prod_{j}2\pi R^{j}\right) 
	\int_{\epsilon^2}^{\infty} \frac{d \tau}{\tau^{6}}\sum_{\boldsymbol{m}\in\boldsymbol{Z}^D}
	e^{-\frac{\pi}{\tau}\sum_{a}\left( m_{a}R^{a}\right)^2} e^{-\frac{\pi}{\tau}\left( \left( 2m_{9}-1\right) R^{9}\right)^2} \nonumber\\
	&
	\times 8\left(8+r+ \sum_{\alpha\in\Delta'_{+}}e^{2\pi i\sum_{a}\left( m_{a}t^{a}\cdot\alpha\right)}e^{2\pi i\left( \left( 2m_{9}-1\right)t^{9} \cdot\alpha\right)}-\sum_{\alpha\in\Delta'_{-}}e^{2\pi i\sum_{a}\left( m_{a}t^{a}\cdot\alpha\right)}e^{2\pi i\left( 2m_{9}-1\right)t^{9} \cdot\alpha}\right),
\end{align}
where we have included the degrees of freedom $8\times 8$ which come from the gravity multiplet.

Let us study the extremal condition of the cosmological constant. From \eqref{potential result}, the first derivatives of $\Lambda^{(10-D)}$ with respect to $t^{j,I}$ are, up to the prefactor which is independent of $t^{j,I}$,
 \begin{align}
 \frac{\partial \Lambda^{(10-D)}}{\partial t^{a,I}}&\sim\sum_{m_{a}\geq 1}m_{a}e^{-\frac{\pi}{\tau}\left( m_{a}R^{a}\right)^2} \sum_{\alpha\in\Delta'}\alpha^{I}\sin\left(2\pi m_{a}t^{a}\cdot \alpha \right) 
 e^{2\pi i\sum_{b\neq a}\left( m_{b}t^{b}\cdot\alpha\right)}
 e^{2\pi i\left( 2m_{9}-1\right)\left(  c+t^{9}\right) \cdot\alpha},\\
 \label{t9derivative}
 \frac{\partial \Lambda^{(10-D)}}{\partial t^{9,I}}&\sim\sum_{m_{9}\geq 1}\left(2m_{9}-1 \right) e^{-\frac{\pi}{\tau}\left( \left(2m_{9}-1 \right)R^{9}\right)^2} \sum_{\alpha\in\Delta'}\alpha^{I}\sin\left(2\pi \left(2m_{9}-1 \right)\left( t^{9}+c\right) \cdot \alpha \right) 
 e^{2\pi i\sum_{a}\left( m_{a}t^{a}\cdot\alpha\right)}.
 \end{align}
If the following condition holds for all $j$, then $t^{j,I}$ is an extremum of $\Lambda^{(10-D)}$:
\begin{align}\label{extremum condition}
t^{j}\cdot\alpha=0~~ (\text{mod 1/2}) ~~\text{for any }\alpha\in \Delta'.
\end{align}
 As in the above example, let us set $t^{a,I}=0$ and focus on the $t^{9,I}$-directions in the moduli space. The condition \eqref{extremum condition} under $t^{a,I}=0$ implies that the number of degrees of freedom for the massless particles after twisted compactification is equal to half of that before the twisted compactification, i.e. $10\times dim(g)$.
For instance, as we can find from Table 1, the configuration \eqref{condition1} with even $q,~q'$ satisfies \eqref{extremum condition} for $\alpha\in\Delta'_{E_8\times E_8}$. Note that \eqref{extremum condition} is a sufficient condition for $t^{j,I}$ to be an extremum. There is a possibility that the first derivatives vanish after summing over $\alpha$. As an simple example, let us consider the case where the gauge group is $U(3)$ with $\alpha^{I}=\left(\underline{+,-,0} \right) \in\Delta'_{U(3)}$ and its twist is given by $c_{I}=\left(0,0,1/2 \right) $. This twist breaks the gauge symmetry to $U(2)\times U(1)$. The parts of the sum of $\alpha$ in \eqref{t9derivative} are
\begin{align}
	\frac{\partial \Lambda^{(10-D)}}{\partial t^{9,1}}&\sim 2\sin \left[ 2\pi \left(2m_{9}-1 \right)\left( t^{9,1}-t^{9,2} \right)\right]   -2\sin \left[ 2\pi \left(2m_{9}-1 \right)\left( t^{9,1}-t^{9,3} \right) \right]\nonumber \\ 
	&=4\cos\left[\pi \left(2m_{9}-1 \right) \left( 2t^{9,1}-t^{9,2}-t^{9,3} \right) \right] \sin\left[\pi \left(2m_{9}-1 \right) \left( t^{9,3}-t^{9,2} \right) \right], \\
	\frac{\partial \Lambda^{(10-D)}}{\partial t^{9,2}}&\sim 2\sin \left[ 2\pi \left(2m_{9}-1 \right)\left( t^{9,2}-t^{9,1} \right)\right]   -2\sin \left[ 2\pi \left(2m_{9}-1 \right)\left( t^{9,2}-t^{9,3} \right) \right] \nonumber\\ 
	&=4\cos\left[\pi \left(2m_{9}-1 \right) \left( 2t^{9,2}-t^{9,1}-t^{9,3} \right) \right] \sin\left[\pi \left(2m_{9}-1 \right) \left( t^{9,3}-t^{9,1} \right) \right], \\
	\frac{\partial \Lambda^{(10-D)}}{\partial t^{9,3}}&\sim-2\sin \left[ 2\pi \left(2m_{9}-1 \right)\left( t^{9,3}-t^{9,1} \right)\right]   -2\sin \left[ 2\pi \left(2m_{9}-1 \right)\left( t^{9,3}-t^{9,2} \right) \right] \nonumber\\ 
	&=-4\cos\left[\pi \left(2m_{9}-1 \right) \left( t^{9,2}-t^{9,1} \right) \right] \sin\left[\pi \left(2m_{9}-1 \right) \left( 2t^{9,3}-t^{9,1}-t^{9,2} \right) \right].
\end{align}
Then, we can find an infinite number of extrema $t^{9,I}=\left(a,a-1/3,a-1/6 \right)$ with $a\in \boldsymbol{R}$, which do not satisfy the condition \eqref{extremum condition}. In \cite{Itoyama:2020ifw}, we have found such extrema in the concrete string models one of which corresponds to the case considered above.

\section*{Acknowledgments}
We thank Nobuhito Maru for insightful remarks to our work. The work of HI is supported in part by JSPS KAKENHI Grant Number 19K03828
and by the Osaka City University (OCU) Strategic Research Grant 2020 for priority area (OCU-SRG2019\_TPR01). The work of SN is supported in part by JSPS KAKENHI Grant Number 21J15497.



\begin{thebibliography}{99}
 	
 	\bibitem{Spontaneous Breaking of Supersymmetry Through Dimensional Reduction}
 	J.~Scherk and J.~H.~Schwarz,
 	Phys.\ Lett.\  {\bf 82B}, 60 (1979);\\
 	R.~Rohm,
 	Nucl.\ Phys.\ B {\bf 237}, 553 (1984);\\
 	C.~Kounnas and B.~Rostand,
 	Nucl.\ Phys.\ B {\bf 341}, 641 (1990).

    \bibitem{Dixon:1986iz}
    L.~J.~Dixon and J.~A.~Harvey,
    Nucl.\ Phys.\ B {\bf 274}, 93 (1986);\\
    L.~Alvarez-Gaume, P.~H.~Ginsparg, G.~W.~Moore and C.~Vafa,
    Phys.\ Lett.\ B {\bf 171}, 155 (1986).
 
    \bibitem{Itoyama:1986ei}
    H.~Itoyama and T.~R.~Taylor,
    Phys.\ Lett.\ B {\bf 186}, 129 (1987); \\
    H.~Itoyama and T.~R.~Taylor,
    FERMILAB-CONF-87-129-T, Proceedings of International Europhysics Conference on High-energy Physics, 25 June-1 July 1987. Uppsala, Sweden (C87-06-25).

    \bibitem{Nair:1986zn}
    V.~P.~Nair, A.~D.~Shapere, A.~Strominger and F.~Wilczek,
    Nucl.\ Phys.\ B {\bf 287}, 402 (1987);\\
    P.~H.~Ginsparg and C.~Vafa,
    Nucl.\ Phys.\ B {\bf 289}, 414 (1987).
 
    \bibitem{Blum:1997gw}
    J.~D.~Blum and K.~R.~Dienes,
    Nucl.\ Phys.\ B {\bf 516}, 83 (1998)
    [hep-th/9707160]; Phys.\ Lett.\ B {\bf 414}, 260 (1997)
    [hep-th/9707148];\\
    S.~Abel, E.~Dudas, D.~Lewis and H.~Partouche,
    JHEP {\bf 1910}, 226 (2019)
    [arXiv:1812.09714 [hep-th]];\\
    H.~Partouche,
    arXiv:1901.02428 [hep-th];\\
    C.~Angelantonj, H.~Partouche and G.~Pradisi,
    arXiv:1912.12062 [hep-th].\\
    S.~Abel, T.~Coudarchet and H.~Partouche,
    arXiv:2003.02545 [hep-th];\\
    T.~Coudarchet and H.~Partouche,
    [arXiv:2011.13725 [hep-th]].

    \bibitem{Itoyama:2019yst}
    H.~Itoyama and S.~Nakajima,
    PTEP {\bf 2019}, no. 12, 123B01 (2019)
    [arXiv:1905.10745 [hep-th]].

    \bibitem{Itoyama:2020ifw}
    H.~Itoyama and S.~Nakajima,
    Nucl. Phys. B \textbf{958}, 115111 (2020)
    [arXiv:2003.11217 [hep-th]].

    \bibitem{Itoyama:2021}
    H.~Itoyama and S.~Nakajima,
    Phys. Lett. B \textbf{816}, 136195 (2021)
    [arXiv:2101.10619 [hep-th]].

    \bibitem{Abel:2015oxa}
    S.~Abel, K.~R.~Dienes and E.~Mavroudi,
    Phys.\ Rev.\ D {\bf 91}, no. 12, 126014 (2015)
    [arXiv:1502.03087 [hep-th]]; Phys.\ Rev.\ D {\bf 97}, no. 12, 126017 (2018)
    [arXiv:1712.06894 [hep-ph]];\\
    B.~Aaronson, S.~Abel and E.~Mavroudi,
    Phys.\ Rev.\ D {\bf 95}, no. 10, 106001 (2017)
    [arXiv:1612.05742 [hep-th]];\\
    S.~Abel and R.~J.~Stewart,
    Phys.\ Rev.\ D {\bf 96}, no. 10, 106013 (2017)
    [arXiv:1701.06629 [hep-th]].
 
    \bibitem{Faraggi:2009xy}
    A.~E.~Faraggi and M.~Tsulaia,
    Phys.\ Lett.\ B \textbf{683}, 314-320 (2010)
    [arXiv:0911.5125 [hep-th]];\\
    J.~M.~Ashfaque, P.~Athanasopoulos, A.~E.~Faraggi and H.~Sonmez,
    Eur.\ Phys.\ J.\ C {\bf 76}, no. 4, 208 (2016)
    [arXiv:1506.03114 [hep-th]];\\
    A.~E.~Faraggi,
    Eur.\ Phys.\ J.\ C {\bf 79}, no. 8, 703 (2019)
    [arXiv:1906.09448 [hep-th]];\\
    A.~E.~Faraggi, V.~G.~Matyas and B.~Percival,
    arXiv:1912.00061 [hep-th];
    Nucl. Phys. B \textbf{961}, 115231 (2020)
    [arXiv:2006.11340 [hep-th]];
    [arXiv:2010.06637 [hep-th]]; [arXiv:2011.04113 [hep-th]];
    [arXiv:2011.12630 [hep-th]];\\
    A.~E.~Faraggi, B.~Percival, S.~Schewe and D.~Wojtczak,
    [arXiv:2101.03227 [hep-th]].
 
    \bibitem{Kounnas:2015yrc}
    C.~Kounnas and H.~Partouche,
    PoS PLANCK {\bf 2015}, 070 (2015)
    [arXiv:1511.02709 [hep-th]]; Nucl.\ Phys.\ B {\bf 913}, 593 (2016)
    [arXiv:1607.01767 [hep-th]]; Nucl.\ Phys.\ B {\bf 919}, 41 (2017)
    [arXiv:1701.00545 [hep-th]];\\
    I.~Florakis and J.~Rizos,
    Nucl.\ Phys.\ B {\bf 913}, 495 (2016)
    [arXiv:1608.04582 [hep-th]];\\
    T.~Coudarchet, C.~Fleming and H.~Partouche,
    Nucl.\ Phys.\ B {\bf 930}, 235 (2018)
    [arXiv:1711.09122 [hep-th]];\\
    T.~Coudarchet and H.~Partouche,
    Nucl.\ Phys.\ B {\bf 933}, 134 (2018)
    [arXiv:1804.00466 [hep-th]];\\
    H.~Partouche,
    Universe {\bf 4}, no. 11, 123 (2018)
    [arXiv:1809.03572 [hep-th]].
    
    \bibitem{Chamseddine:1980cp}
    A.~H.~Chamseddine,
    Nucl. Phys. B \textbf{185}, 403 (1981).
    
    \bibitem{Chapline:1982ww}
    G.~F.~Chapline and N.~S.~Manton,
    Phys. Lett. B \textbf{120}, 105-109 (1983).
 
    \bibitem{Green:1984sg}
    M.~B.~Green and J.~H.~Schwarz,
    Phys. Lett. B \textbf{149}, 117-122 (1984).
 
	\bibitem{Candelas:1985en}
	P.~Candelas, G.~T.~Horowitz, A.~Strominger and E.~Witten,
	Nucl. Phys. B \textbf{258}, 46-74 (1985).

	\bibitem{Manton:1979kb}
	N.~S.~Manton,
	Nucl. Phys. B \textbf{158}, 141-153 (1979).

	\bibitem{Fairlie:1979at}
	D.~B.~Fairlie,
	Phys. Lett. B \textbf{82}, 97-100 (1979).
	
	\bibitem{Hosotani:1983xw}
	Y.~Hosotani,
	Phys. Lett. B \textbf{126}, 309-313 (1983).

	\bibitem{Hosotani:1988bm}
	Y.~Hosotani,
	Annals Phys. \textbf{190}, 233 (1989).

	\bibitem{Hatanaka:1998yp}
	H.~Hatanaka, T.~Inami and C.~S.~Lim,
	Mod. Phys. Lett. A \textbf{13}, 2601-2612 (1998)
	[arXiv:hep-th/9805067 [hep-th]].
	
	\bibitem{Quiros:2003}
	M.~Quiros,
	[arXiv:hep-ph/0302189 [hep-ph]].
 \end{thebibliography}
\end{document}